\documentclass[prb,twocolumn,showpacs,showkeys,amsmath,amssymb]{revtex4}
\usepackage{graphicx}
\usepackage{dcolumn}
\usepackage{bm}

\begin{document}

\vskip 3cm
\title{
Cyclotron Resonance of Wigner Crystals on  Liquid Helium
}

\author{Van An Dinh$^{1,2,3}$}
\email{divan@cmp.sanken.osaka-u.ac.jp}
\author{Saitoh Motohiko$^1$}
 \affiliation{%
$^1$Department of Physics,~Graduate School of Science,~Osaka University,\\ 
1-1 Machikaneyama, Toyonaka 560-0043, Japan.\\
$^2$Department of Condensed Matter Physics, and $^3$Department of Computational Nanomaterials Design,\\ Institute of Scientific and Industrial Research, Osaka University,\\ 8-1 Mihogaoka, Ibaraki, Osaka 567-0047, Japan}

\begin{abstract} 
The cyclotron resonance of the  correlated two-dimentional electrons on liquid helium in high magnetic fields is investigated on the basis of the newly developed theory. Electrons are assumed to form a Wigner crystal, and the electron correlation effect  is taken into account through the self-consistently determined Wigner phonons, where the electron-electron interaction, the electron-ripplon and electron-vapor atom scatterings are considered on the same footing. The numerical calculations  show a very good agreement with  experiments without any fitting parameters.\end{abstract}

\maketitle

\section{Introduction}
  The cyclotron resonance (CR) is considered to be  one of the basic techniques for the  investigation of the properties of the 2D electrons, particularly the electrons which form a Wigner crystal~\cite{Grimes,Fisher}.  Though there are cumulative theories on the CR of the Wigner crystal on liquid helium~\cite{Dykman,Monarkha}, this problem is still controversial as to how the electron Coulomb interaction is taken into account. In this report, we present a new way to incorporate the electron correlation and the electron scattering effects, which was successful to interpret the CR of 2D electrons in the semiconductor heterostructures~\cite{An}, and give the theoretical analysis of the CR of the Wigner crystal on the surface of liquid helium in  high magnetic fields. The  electron-electron interaction,  electron-ripplon and electron-vapor atom scattering are taken into account through the Wigner phonon spectra modified by the sccatterings. The present systematic method should be compared to the existing phenomenological theories~\cite{Monarkha}. The dynamic structure factor  and the CR linewidth are calculated, and the good agreement is obtained with experiments without any phenomenological parameters.
  
\section{\label{model}Model}
We consider that the 2D electrons  on the surface of liquid helium form a Wigner crystal (WC) due to the electron Coulomb interaction. Then the Hamiltonian of the system is given by
\begin{equation}
H=\sum_{{\bf k},j=\pm}{\hbar\Omega^{(0)}_{{\bf k}j}a^\dagger_{{\bf k}j}a_{{\bf k}j}}
+V_{\rm e-r}+V_{\rm e-v},
\end{equation} 
where 
$\Omega^{(0)}_{{\bf k}j}$'s are the unperturbed Wigner phonon frequency in a magnetic field, $a^\dagger_{{\bf k}j}$ and $a_{{\bf k}j}$ their creation and annihilation operators, respectively,  $j\  (j=\pm)$ denotes  the phonon mode, and $V_{\rm e-r}$ and $V_{\rm e-v}$ are the scattering potentials due to ripplons and vapor atoms of $^4$He, respectively. 

The ripplon scattering potential is expressed as
\begin{equation}
V_{\rm e-r}=\frac{1}{\sqrt{A}}\sum_{n,\bf q}{eF(q)\sqrt{\frac{\hbar q}{2\rho\nu_q}}e^{i{\bf q}\cdot{\bf r}_n}(b_{\bf q}+b_{\bf q}^\dagger)},
\end{equation}
where  $A$ is the area of the system,  $\nu_q=(\alpha/\rho)^{1/2}q^{3/2}$  the ripplon frequency with $\alpha$ and $\rho$  the surface tension constant and mass density of liquid helium, respectively, $b_{\bf q}^\dagger$ and $b_{\bf q}$  are the creation and annihilation operators of a ripplon with wave vector $\bf q$,  respectively,  ${\bf r}_n$  the position vector of the $n$th electron parallel to the surface, and $F(q)$  the $q$-dependent effective holding field which takes the  form:
\begin{equation}
eF(q)=eE_z+\frac{\hbar^2q^2}{2ma_0}w\left(\frac{qb}{2}\right).
\end{equation}
Here $E_z$ is the applied holding electric field perpendicular to the liquid surface, $a_0$ and $b$ are given by~\cite{Saitoh}
\begin{eqnarray}
 a_0&=&\frac{4\hbar^2(\epsilon_{\rm He}+1)}{me^2(\epsilon_{\rm He}-1)},\\
b&=&\!\frac{4a_0}{3\lambda}\!\sinh{\left(
\frac{1}{3}\sinh^{-1}{\frac{9\lambda}{4}}
\right)},\\ 
\lambda&=&\sqrt{2ma_0^3eE_z}/\hbar,
\end{eqnarray}
where $\epsilon_{\rm He}$ is the dielectric constant of  liquid helium, and the function $w(x)$ is given by
\begin{equation}
w(x)\!\!=
\!\!\left\{
	\begin{array}{ll}
\!\!\frac{1}{\!1-\!x^2}[\frac{1}{\sqrt{\!1-\!x^2}}\!\ln{\!\frac{1+\sqrt{1-x^2}}{x}}-1], (x<1),\\
\!\!\frac{1}{\!x^2-\!1}[1-\frac{1}{\sqrt{\!x^2-\!1}}\!\arctan\!{\sqrt{x^2-1}}], (x>1).
	\end{array}
	\right.
\end{equation}
 
 The vapor scattering potential  $V_{\rm e-v}$ is a contact type interaction and given by~\cite{Saitoh}
\begin{equation}
 V_{\rm e-v}=V_0\sum_{a,n}{\delta({\bf R}_a-{\bf r}_n)\delta(Z_a-z_n)},
\end{equation}
where $({\bf R}_a,Z_a)$ is the position vector of the $a$th vapor atom and 
$ z_n$ the perpendicular component of the position of the $n$th electron, and $V_0$ is the effective interation strength which is related to the cross section $A_{\rm He}$ of a helium atom by 
$V_0^2=\pi\hbar^4A_{\rm He}/m^2$. 

 The Wigner phonons in  a magnetic field $B$ are related to the phonons without magnetic fields by
\begin{equation}\label{eq:Omega}
\Omega^{(0)}_{{\bf k}\pm}\!\!=\!\!\frac{\sqrt{\!\omega_c^2+(\!\omega^{(0)}_{{\bf k}l}+\omega^{(0)}_{{\bf k}t}\!)^2}\pm \sqrt{\!\omega_c^2+(\!\omega^{(0)}_{{\bf k}l}-\omega^{(0)}_{{\bf k}t}\!)^2}}{2}, 
\end{equation}
where $\omega^{(0)}_{{\bf k\lambda}} (\lambda=l,t)$ are frequencies  corresponding to the longitudinal ($l$) and the transverse ($t$)  modes of the Wigner phonon without magnetic fields,
 and $\omega_c=eB/mc$ is the cyclotron frequency. 
When the scattering potentials are present, $\omega^{(0)}_{{\bf k\lambda}}$ will be modified to $\omega_{{\bf k\lambda}}$ by the scattering and they 
 should be determined self-consistently. It follows that the spectra 
 in such a case have a gap  approximately given by~\cite{An}
 \begin{equation}
 \omega_{{\bf k}\lambda}^2=\omega_{{\bf k}\lambda}^{(0)2}+v^2,
 \end{equation}
 \begin{equation}
v^2\approx\frac{1}{A}\sum_{\bf q}{\frac{q^2U_q}{2mT}\exp{\left(-\frac{q^2}{4m\omega_c}\right)}},
 \end{equation}
 where $T$ is the temperature in  units of energy, and $U_q$  the coupling function given by
\begin{equation}
\label{eq:u}U_q=\frac{Te^2F^2(q)}{\alpha q^2}+\frac{\hbar^3}{m\tau_{\rm v}}.
\end{equation}
The first term in Eq.~(\ref{eq:u}) corresponds to the electron-ripplon scattering, and the second to the electron-vapor atom scattering where $\tau_{\rm v}=8mb/3\pi A_{\rm He}n_{\rm v}$ is
the collision time due to  helium vapor atoms with $n_{\rm v}$ being 
 the vapor density.  
In accordance with this, $\Omega^{(0)}_{{\bf k}j}$ are modified to $\Omega_{{\bf k}j}$ in the same manner as in Eq.~(\ref{eq:Omega}).
\section{\label{dsf}Dynamic structure factor}
As usual, the dynamic structure factor (DSF) is defined through the density-density correlation function $S_{nm}(q,t)=<e^{i{\bf q\cdot u}_n(t)}e^{-i{\bf q\cdot u}_m(0)}>$ by
\begin{equation}
S_{n-m}(q,\omega)=\displaystyle\int\limits_{-\infty}^{\infty}e^{i\omega t}S_{nm}(q,t)dt,\end{equation}
where ${\bf u}_n$ is the displacement vector of the electron at the $n$th-site, and $<\!~\cdots\!~>$ denotes the thermal average.
This can be  calculated readily by using the expression of the Fourier transform of the displacement vector ${\bf u}_n$ in the quantized representation~\cite{An}.

 In a high magnetic field ($\hbar\omega_c\gg T$)
  the dominant DSF with $n=m$ can be expressed as a sum over  contributions from all the Landau levels~\cite{An}:
 \begin{eqnarray}\label{eq:sqf}
S_{0}(q,\omega)\!&=&\!\frac{2\sqrt{\pi}\hbar}{\Gamma}\sum_{n=0}^{\infty}{\frac{x^{n-\frac{1}{2}}}{n!}e^{\hbar(\omega-n\omega_c)/2T}}\\
\!\!&\times&\!\exp{\!\left[-x\left(\!1+\!\left(\frac{\Gamma}{4T}\!\right)^2\right)\!-\frac{\hbar^2(\omega-n\omega_c)^2}{x\Gamma^2}\right]}.\nonumber
\end{eqnarray} 
Here, $x=\hbar q^2/2m\omega_c$, and the broadening parameter,
 $\Gamma$, is defined by
\begin{equation} 
\Gamma^2=\frac{2\hbar T}{\omega_cN}\sum_{{\bf k},\lambda}{\omega^2_{{\bf k}\lambda}}={\Gamma_{\rm e}^2+\Gamma_{\rm r}^2+\Gamma_{\rm v}^2},
\end{equation}
 where $\Gamma_{\rm e}$ is related to the the electron Coulomb interaction, $\Gamma_{\rm r}$ to the electron-ripplon,  and $\Gamma_{\rm v}$ to electron-vapor atom  scatterings, respectively, and are given  by
\begin{eqnarray}
 \Gamma_{\rm e}&=&\eta\omega_0\sqrt{\!\frac{\hbar T}{\omega_c}},\\
\Gamma_{\rm r}&=&\sqrt{\frac{T}{\pi\alpha}}eF(q\!=\!\frac{\sqrt{2}}{l_c}),\\ 
\Gamma_{\rm v}&=&\hbar\sqrt{\frac{2\omega_c}{\pi\tau_{\rm v}}},
\end{eqnarray}
where $\eta$ is a numerical constant which may depend weakly on the plasma parameter~\cite{Fozooni} and $\approx 1.21$ at zero temperature, $l_c=(c\hbar/eB)^{1/2}$ the Landau magnetic length, and $\omega^2_0=4(\pi n_e)^{3/2}e^2/m$ with $n_e$ the surface density of electrons.
 \begin{figure}[h]
\begin{center}\leavevmode 
\includegraphics[width=1.0\linewidth,height=1.0\linewidth]{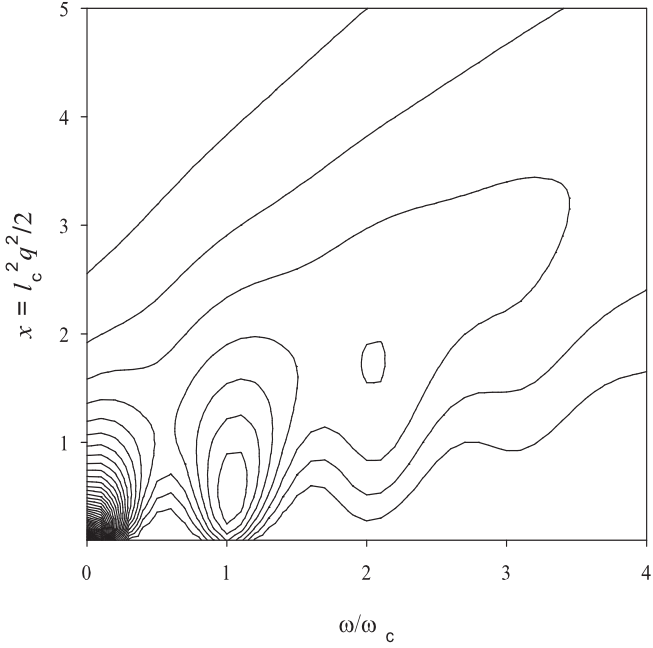}
\caption{\label{fig:dsf} Contour map of $S_{0}(q,\omega)$ 
 at $B=0.7$~T, $T=0.72$~K and $n_e~\!=~\!2\cdot~10^{8}$~cm$^{-2}$.}
\end{center}
\end{figure}

The present result Eq.~(\ref{eq:sqf}) differs appreciably from the results by Monarkha {\it et al.}~\cite{Monarkha} who determined the DSF phenomenologically, in which the factor $\exp{[-x(\Gamma/4T)^2]}\times\exp{[\hbar(\omega-n\omega_c)/2T]}$ is missing, and so the functional form with respect to $x$ is different from ours. 
This difference has lead to an overestimation of the contributions from the higher Landau levels in Ref. ~4.  
Also, note that the broadening parameter $\Gamma$ contains the contributions of the scattering effects and so, as will be shown in the next section, the linewith  goes over  without divergence to the self-consistent Born approximation for independent electrons~\cite{SaitohCR},  when the electron correlation effect becomes negligible. Therefore, this is regarded as an improvement over Dykman's result~\cite{Dykman} in which the linewidth diverges as $n_{ e}$ goes to zero.

Figure 1 illustrates an example of the DSF.  As seen in Fig.~1, the DSF has  local peaks at ($\omega/\omega_c\approx n+(n-{1}/{2}){\Gamma^2}/{4T\hbar\omega_c},x\approx n-{1}/{2}$) for $n\neq 0$. The most important contributions to the DSF come from the terms with $n=0$, and $ 1$, and  small enough $q$.   

\section{\label{crlw}Cyclotron Resonance Linewidth}
 The width function of the absorption line-shape is defined through the imaginary part of the memory function of the conductivity. Similar to the case of electrons in semiconductor heterostructures~\cite{An}, the width function of the present case takes the following form: 
 \begin{equation}
\label{eq:g}\gamma(\omega)=\frac{1-e^{-\hbar\omega/T}}{\hbar\omega}\frac{1}{A}\sum_{\bf q}{\frac{ q^2U_q}{4m}S_{0}(q,\omega)}.
\end{equation}
Inserting $S_{0}(q,\omega)$ in Eq.~(\ref{eq:sqf}) into Eq.~(\ref{eq:g}), we can obtain the final expression for the width function. At resonance ($\omega=\omega_{\rm c}$), the CR linewidth is given by
\begin{equation}
\gamma(\omega_c)=\gamma_{\rm r}(\omega_c)+\gamma_{\rm v}(\omega_c).
\end{equation}
In the above, the CR linewidth $\gamma_{\rm r}(\omega_c)$ due to the ripplon scattering  has the explicit form
\begin{equation}
\label{eq:gr}\!\!\gamma_{\rm r}(\omega_c)\!=\!\frac{T(eE_z)^2}{2\sqrt{\pi}\hbar\Gamma\alpha}\sinh{\!\frac{\hbar\omega_c}{2T}}\!\sum_{n=0}^\infty{\!\frac{e^{-n\hbar\omega_c/2T}}{n!}G_n},
\end{equation}
where $G_n$ is defined by
\begin{eqnarray}
G_n\!&=&\int_0^\infty\!\!{x^{n-\frac{1}{2}}V^2(x)\!
e^{-x(1+(\frac{\Gamma}{4T})^2)-\frac{\hbar^2\omega_c^2(1-n)^2}
{x\Gamma^2}}\!dx},\\
V(\!&x&\!)=1+\frac{\hbar^2}{2ma_0l_{c}^2eE_z}xw\left(\frac{b}{l_c}\sqrt{\frac{x}{2}}\right).
\end{eqnarray}
\noindent 
The CR linewidth $\gamma_{\rm v}(\omega_c)$ due to the vapor atom scattering  is given by
\begin{eqnarray}
\label{eq:gv}
\gamma_{\rm v}(\omega_c)&=&\frac{2\hbar\omega_{c}}{\sqrt{\pi}\Gamma\tau_{\rm v}}\sinh{\frac{\hbar\omega_c}{2T}}\sum_{n=0}^\infty{\frac{e^{-n\hbar\omega_c/2T}}{n!}}\nonumber\\
&\times&\!\!\left(\!\frac{\hbar\omega_c\!\mid\! 1 -n\mid}{\xi\Gamma}\!\!\right)^{\!n+\frac{3}{2}}\!\!\!\!\!K_{n+\frac{3}{2}}\!\!\left(\!\frac{2\xi\hbar\omega_c\!\mid\!1-n\!\mid}{\Gamma}\!\right)\!,
\end{eqnarray}
where $\xi=\sqrt{1+(\Gamma/4T)^2}$ and $K_\nu(z)$ is the modified Bessel function of the second kind.

\begin{figure}[h]
\begin{center}\leavevmode
\includegraphics[width=1.0\linewidth, height=1.2\linewidth]{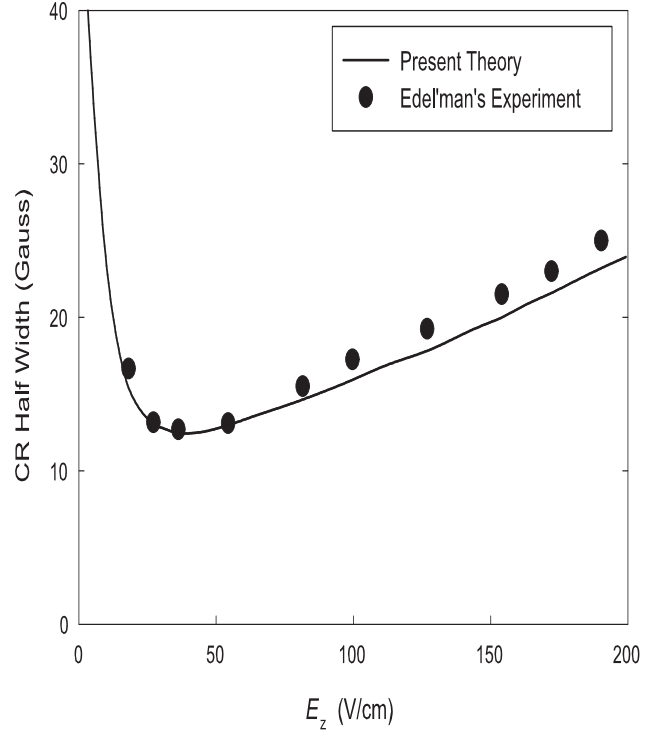}
\caption{The CR linewidth {\it vs.} holding electric fields $E_z$ at $T=0.72$~K. The holding electric field is taken to be the saturation value  $E_z=2\pi n_e$.  Dots are taken from the experiments by Edel'man~\cite{Edelman}.}
\end{center}
\end{figure}
Figure 2 illustrates the applied electric field dependence of the CR
 linewidth  in the condition that the electron density is saturated: $E_z=2\pi n_e$. The solid curve represents the results of the present theory, and the dots  the experimental results by Edel'man~\cite{Edelman}.
A comment is in order here.  The existence of the Wigner crystal phase is marginal in Edel'man's experiment.  However, our theory does work beyond the Wigner crystal phase and is applicable even in such a case, since the electron correlation effect taken into account through $\Gamma_{\rm e}$ has the same form as in Ref.~\cite{Dykman,Fozooni} though with a different physical 
interpretation.
At this temperature both of ripplon and vapor scatterings are equally 
important.
As shown in Fig.~2, the CR linewidth first decreases with increasing $E_z$ in the small  electric field range ($E_z<30$~V/cm), and then increases for the larger holding fields. It is emphasized here that the term with $n=1$ in the CR linewidth in Eq.~(\ref{eq:gr}) and Eq.~(\ref{eq:gv}) monotonocally decreases with the electron density, and therefore the dip of the CR linewidth cannot be reproduced without  the terms with $n=0$ and $n\geq 2$. 
Our numerical results of the CR linewidth  agree quantitatively  with the experiment and better than  the previous theories. Note that we have not introduced any fitting prameters.
 \section{Conclusion}
 We have presented  a theoretical analysis of the cyclotron resonance of the Wigner crystal on the surface of liquid helium under quantizing magnetic fields, not phenomenologically as was done in previous theories, but by taking the electron correlation and the electron scatterings   on the same footing. We studied the DSF and the CR linewidth by fully considering all the contributions from the higher Landau levels.  The numerical calculations show that the contributions from higher Landau levels are important to reproduce  the non-monotonic dependence in $E_z$ of the CR linewidth properly. 
 The electron density dependence of the CR linewidth is shown to be in the quantitative agreement with experimental results by Edel'man~\cite{Edelman} without any fitting parameters.


\noindent
\begin{thebibliography}{9}
\bibitem{Grimes}C. C. Grimes and G. Adams, Phys. Rev. Lett. {\bf 42}, 795 (1979).
\bibitem{Fisher}D.S. Fisher, B.I. Halperin and P.M. Platzman, Phys. Rev. Lett. {42}, 798 (1979).
\bibitem{Dykman}M. I. Dykman, J. Phys. C {\bf 15}, 7397 (1982).
\bibitem{Monarkha} Yu. P. Monarkha, E. Teske and P. Wyder, Phys. Reports {\bf 370}, 1 (2002).
\bibitem{An} Van An Dinh and M. Saitoh, Physica E {\bf 18}, 155 (2003).\\
Van An Dinh and M. Saitoh,  J. Phys. Soc. Jpn. {\bf 70}, No. 7 (2003).
\bibitem{Saitoh}M. Saitoh, J. Phys. Soc. Jpn. {\bf 42}, 201 (1977).
\bibitem{Fozooni} P. Fozooni, P.J. Richardson, M.J. Lea, M.I. Dykman, C. Feng-Yen and A. Blackburn, J. Phys.:Condensed Matter {\bf 8}, L215 (1996).
\bibitem{SaitohCR} M. Saitoh, J. Phys. {\bf C16}, 6983 (1983).
\bibitem{Edelman} V. S. Edel'man, Sov. Phys. JETP {\bf 50}, 338
 (1979).
 
\end{thebibliography}
\end{document}